\documentclass[a4paper,twoside]{article}

\usepackage{epsfig}
\usepackage{subcaption}
\usepackage{calc}
\usepackage{amssymb}
\usepackage{amstext}
\usepackage{amsmath}
\usepackage{amsthm}
\usepackage{multicol}
\usepackage{pslatex}
\usepackage{apalike}
\usepackage{algorithm2e}
\usepackage{pifont}
\usepackage{tabularx}
\usepackage[bottom]{footmisc}
\usepackage{url}
\usepackage{SCITEPRESS}

\usepackage{xcolor}
\usepackage{listings}
\usepackage{pifont}
\usepackage{float}

\begin{document}

\title{An Analysis of Modern Web Security Vulnerabilities Inside WebAssembly Applications}

\author{\authorname{Lorenzo Corrias\sup{1}\orcidAuthor{0009-0004-0550-087X}, Lorenzo Pisu\sup{1}\orcidAuthor{0009-0001-0129-1976} Davide Maiorca\sup{1}\orcidAuthor{0000-0003-2640-4663} and Giorgio Giacinto\sup{1}\sup{2}\orcidAuthor{0000-0002-5759-3017}}
\affiliation{\sup{1}Department of Electronic and Computer Engineering, University of Cagliari, Italy\\ \sup{2}National Interuniversity Consortium for Informatics, Rome, Italy}
\email{l.corrias32@studenti.unica.it, \{lorenzo.pisu, davide.maiorca, giacinto\}@unica.it}}


\abstract{The growth in the adoption of WebAssembly (WASM) has given rise to a rapidly increasing landscape of binary applications that are natively ported to the environment of websites. The flexibility of WASM has made it the preferred way to run fast and resource-heavy applications, replacing a field that JavaScript previously monopolized. Despite its success, researchers have raised concerns over the security implementations of WASM, demonstrating that binary vulnerabilities, such as Buffer Overflows and Use After Free, remain a present danger for WASM binaries. Our work aims to demonstrate that such vulnerabilities, when occurring on a WebAssembly module, can affect the behavior of a web application in unexpected ways, enabling an attacker to exploit typical web security flaws. We provide several scenarios as examples of how each binary vulnerability might lead to a web security one, such as SQL Injections, XS-Leaks, and SSTIs. Our results show that binary vulnerabilities can invalidate common security mechanisms, demonstrating how the safety of WASM modules remains a problem that needs to be addressed. We also provide a list of best practices and defensive strategies that developers can implement to mitigate the risks associated with running unsafe WASM modules in their web applications.}

\onecolumn \maketitle \normalsize \setcounter{footnote}{0} \vfill

\section{Introduction}
\label{sect:introduction}

Modern web development increasingly requires complex applications running natively on the web, driving adoption of fast, efficient technologies like WebAssembly (WASM), now used by sites such as Figma and Google Earth.

Despite its strengths, WASM lacks common low-level protections (stack canaries, ASLR, safe unlinking) \cite{lehmann2020everything,massidda2024bringing,mcfadden2018security}, leaving buffer overflows and use-after-free exploitable for control-flow hijacking \cite{lehmann2020everything}. While the WASM VM blocks some attacks (e.g., shellcode), we show that these flaws can still cause classic web exploits, compromising the host.

Although WASM’s security model has been scrutinized over the years, little work maps low-level WASM bugs to web-layer attacks. Web security often targets high-level code, overlooking the impact of integrating compiled modules.

We show how unsafe WASM can introduce atypical web flaws that evade traditional analysis. We build PoC web services with intentionally vulnerable WASM modules to demonstrate how memory-safety bugs become high-impact attacks, including SQL injection and XS-Leaks.

For reproducibility, each PoC offers step-by-step instructions or a Python demo script. Modules are written in C and compiled with Emscripten to showcase classic exploitation; findings generalize to other unsafe languages (e.g., C++).

The primary contributions of this work are:

\begin{itemize}
    \item We analyze memory/control-flow bugs (stack buffer overflows, use-after-free, integer overflows, format strings) and show how they enable web exploits (SQLi, SSTI, XS-Leaks), chosen for their historical weight (SQLi) or novelty in WASM (SSTI, XS-Leaks).
    \item We propose a reproducible methodology linking compiled-code flaws to web-layer impact, combining vulnerability selection, PoC building, exploit automation, and impact assessment.
    \item We create reproducible PoCs embedding vulnerable WASM modules that realize concrete exploits (e.g., SQL injection via buffer-overflowed queries), with minimal Node.js server/client code and Python scripts for independent checks.
    \item We publish all PoCs, scripts, and docker containers for full reproducibility \footnote{https://anonymous.4open.science/r/wasm-pocs/}.
    \item We highlight under-appreciated WASM attack surfaces, quantify impact, and explain how memory layout and imports/exports affect web security, giving actionable guidance on safer compilation, runtime hardening, JS-WASM boundary checks, and deployment configuration.
\end{itemize}

These contributions clarify WASM security risks, provide reusable artifacts, and offer guidance to mitigate threats from low-level compiled code in web runtimes.

section{Background}
\label{ch:background}

WebAssembly provides a binary compilation format for languages like C, C++, and Rust, enabling near-native execution speed on the web alongside JavaScript. Maintained by the World Wide Web Consortium \cite{w3cwasm} since 2015, the WebAssembly standard defines an instruction set architecture (ISA) similar to assembly language \cite{webassemblyspec}. Instructions are binary encoded (.wasm) with a human-readable text format (.wat) also available. A Virtual Machine executes WASM programs, providing faster execution than interpreted languages while maintaining browser isolation. The WASM specification \cite{w3cwasmjs} defines an interface for JavaScript communication, allowing independent environments to exchange information.

WASM enables execution of computationally expensive code without performance compromise. Recent works \cite{10277917,9678831} show WASM outperforms JavaScript even for simple tasks. WASM modules are generated by compilers translating code from various languages, including compiled languages via JIT/AOT compilers in browsers \cite{webassemblysupport} and interpreted languages (e.g., Python \cite{pyodide}) via interpreters \cite{wasm3}. External compilers also enable server-side execution \cite{wasmer}\cite{wasmtime}. This portability makes WebAssembly suitable for migrating legacy applications to the web without code restructuring. WASM can also execute outside browsers using technologies like WASI \cite{wasi}, which provides a POSIX interface to limit program capabilities.

Our work focuses on how low-level vulnerabilities in WASM modules can lead to the exploitation of web security vulnerabilities. In particular, we decided to analyze vulnerabilities that are common in web applications, following the methodology outlined in Chapter \ref{subsec:meth_web}. The vulnerabilities we selected for our analysis are:

\begin{itemize}
    \item SQL Injection (SQLi). Known since the late 1990s, SQLi is among the top 3 most exploited web vulnerabilities according to OWASP TOP10 2021 \cite{sqlowasp}. It occurs when unsanitized user input is incorporated into database queries, allowing attackers to alter query semantics and cause unintended behavior. Common defenses include prepared statements and input sanitization \cite{preparedstat}.
    \item Server-Side Template Injection (SSTI). SSTI occurs when user input is unsafely embedded into server-side templates, enabling arbitrary code execution within the server's context. This can lead to data leakage and remote code execution, making it more dangerous than SQLi \cite{pisu2024survey}.
    \item Cross-Site Leaks (XS-Leaks). XS-Leaks exploit browser side-channels to leak information between web pages, potentially circumventing browser defenses like Same-Origin Policy (SOP) \cite{sop} and Cross-Origin-Opener-Policy (COOP) \cite{coop}. Our work focuses on timing attacks where adversaries force heavy computation and measure delays to infer secret-dependent behavior.
  \end{itemize}

\section{Related Works}
\label{sec:related_works}

Several works have explored exploiting binary vulnerabilities in WASM modules. Our work builds on three key publications.

\par Mcfadden et al. \cite{mcfadden2018security} pioneered binary security analysis of WASM modules compiled from C, providing initial proof-of-concept exploits demonstrating attack possibilities and limitations.
\par Lehmann et al. \cite{lehmann2020everything} expanded this work, identifying exploitable binary errors in WASM modules and highlighting the absence of standard C executable defenses that could prevent certain exploits.
\par Massidda et al. \cite{massidda2024bringing} identified new exploitation methods for vulnerabilities previously deemed unexploitable \cite{mcfadden2018security}.
\par We extend these works by demonstrating how binary vulnerabilities in WASM modules can be leveraged to exploit web security vulnerabilities like SQL Injections and SSTIs in realistic applications. This work shows that C vulnerabilities threaten WASM applications not only through their traditional destructive effects (e.g., arbitrary code execution), but also through unexpected interactions between vulnerable WASM modules and their host web applications.

\section{Methodology}
\label{sec:methodology}

This section outlines our methodology regarding the selection of the binary and web security vulnerabilities that we investigated.

\subsection{Binary Vulnerabilities}
\label{subsec:meth_binary}

The PoCs formulated in this research rely on binary vulnerabilities to compromise the underlying WASM module. Drawing heavily on prior research by Massidda et al. \cite{massidda2024bringing}, McFadden et al. \cite{mcfadden2018security}, and Lehmann et al. \cite{lehmann2020everything}, we categorized potential vulnerabilities based on the primitives they provide to an attacker:

\begin{itemize}
    \item \textbf{Arbitrary Write:} Vulnerabilities allowing an adversary to write data into the program's memory.
    \item \textbf{Arbitrary Read:} Vulnerabilities allowing an adversary to read data from the program's memory.
\end{itemize}

Based on these categories, we selected four specific vulnerabilities to investigate:

\textbf{Stack-based Buffer Overflow} (Arbitrary Write),
\textbf{Use After Free} (Arbitrary Write),
\textbf{Integer Overflow} (Arbitrary Write),
and \textbf{Uncontrolled Format String} (Arbitrary Read and Write).

These were chosen for their prevalence and their direct applicability to WASM environments. While we leverage payloads derived from Massidda et al. \cite{massidda2024bringing}, we also introduce advancements in exploitation techniques, particularly regarding format strings.

Conversely, certain vulnerabilities were deemed out of scope. Improper array index validation, while exploitable, was excluded to prioritize more pervasive memory corruption issues. Similarly, redirecting indirect calls, a vulnerability that allows overwriting function pointers to alter a program's control flow, was excluded, as it was not considered sufficiently correlated to the specific web security vulnerabilities analyzed in this research.

\subsection{Web Security Vulnerabilities}
\label{subsec:meth_web}

To analyze web security vulnerabilities within the context of WASM modules, we chose to classify them into two distinct categories:

\begin{itemize}
    \item \textbf{Direct or Chained:} "Chained" vulnerabilities occur when the output of a compromised WASM module is incorporated into the web application's execution flow. In this case, the exploitation consists of two phases: compromising the WASM module via a binary vulnerability, followed by the web application processing the polluted output. "Direct" vulnerabilities, in contrast, are contained entirely within the WASM module, meaning that the dependent web application does not interact with the polluted data.
    \item \textbf{Blind or Not-blind:} "Not-blind" vulnerabilities return output confirming the attack's success (e.g., leaked data). "Blind" vulnerabilities return no output, requiring side-channel analysis (such as timing attacks) to infer the application's state.
\end{itemize}

Using this classification, we decided to investigate these three vulnerabilities:

\begin{itemize}
    \item \textbf{SQL Injection (Direct, Not-blind):} Chosen due to the prevalence of WASM libraries for SQL management \cite{sqljs} \cite{sqlitewasm}. We simulate an environment where binary vulnerabilities allow query tampering.
    \item \textbf{Server-Side Template Injection (SSTI) (Chained, Not-blind):} This vulnerability was chosen because it demonstrates the risk of treating WASM output as trusted. We showcase how a binary vulnerability can pollute WASM output, which is subsequently rendered by a template engine to execute harmful behaviors.
    \item \textbf{XS-Leaks (Direct, Blind):} Selected to represent the most difficult exploitation category. We describe a PoC enabling a Regular Expression Denial-of-Service (ReDoS) \cite{redos}, where timing differences allow an attacker to infer sensitive information, such as user secrets, despite the isolation of the WASM VM.
\end{itemize}

We explicitly exclude Prototype Pollution and DOM Clobbering, as WASM modules are generally disallowed from direct interaction with the DOM or the JavaScript global object, making these vectors unrealistic in this specific context. Race conditions were also excluded due to the complexity of implementing concurrent execution exploits in the C-to-WASM translation layer.

\section{Experimental Results}
\label{ch:expresults}

This chapter details the developed Proof of Concepts (PoCs) and experimental results regarding WebAssembly module exploitation. The explanation of the PoCs is structured around the three previously defined web security vulnerabilities in section \ref{subsec:meth_web}. The analysis examines the specific binary vulnerabilities facilitating each attack vector. For some sections, we provide code samples to demonstrate the identification and remediation of undefined behaviors, accompanied by architectural recommendations for prevention. The chapter concludes with a discussion of the broader implications of these findings.

\subsection{SQL Injections}
\label{sec:resultssqli}

To investigate how binary vulnerabilities in WASM modules can escalate to SQL Injection (SQLi) attacks, we designed an architecture consisting of a NodeJS/Express frontend acting as an interface for a backend WASM module compiled from C, which manages an SQLite database \cite{sqlitewasm}. The C program that queries the database uses parametrized statements, guarding it against SQL injection attacks. However, we demonstrate that binary exploitation can bypass these protections. Exploitation relies on two primary primitives:

\begin{itemize}
    \item If the attacker has access to a \textbf{write primitive}, it is able to overwrite the memory location containing the parametrized query string before execution.
    \item If the attacker has access to a \textbf{read primitive}, it is able to leak sensitive data returned by a query, bypassing frontend filters.
\end{itemize}

\subsubsection{Stack-based Buffer Overflow}
\label{subsubsec:bof_sqli}

This vulnerability provides an attacker with a write primitive. Even when using parametrized statements, an application remains vulnerable if the query string itself is stored on the stack and can be corrupted by an adjacent buffer overflow.

For example, consider the following buffer overflow payload: \verb|"A"*32 + "SELECT 1"|. If an attacker is able to inject this input into a part of the program that is vulnerable to a buffer overflow, it can fill a variable's buffer and overwrite the variable containing the original SQL query, replacing it with a malicious one. Successful exploitation, however, is constrained by a number of different factors. These are:

\begin{itemize}
    \item The stack layout. In order to achieve an SQLi, the target string containing the query must reside at a higher memory address than the variable being overridden by the overflow.
    \item The scope. To have a successful override, the target and buffer must share the same scope or stack frame as the one being overridden.
    \item The number of binding parameters in the query. Due to the fact that, typically, prepared statements contain a number of binding parameters, which are then replaced by the actual variables once the query is executed (they are usually represented by a "\verb|?|" character). An attacker must make sure to override the query with one that has the same number of parameters, so that the execution doesn't error out.
\end{itemize}

Despite the numerous limitations, buffer overflows in this context can be exploited by an attacker with relative ease. Our researches demonstrate how the use of prepared statements, which is the standard defence used to guard against SQL injections, is ineffective if the query can be overridden entirely.

\subsubsection{Uncontrolled Format String}
\label{subsubsec:ufs_sqli}

Uncontrolled format strings enable both read and write primitives. However, their unpredictable nature—tampering with static memory segments—poses significant challenges for SQLi exploitation, particularly regarding payload size and memory layout. We distinguish two scenarios based on available primitives.
\par For write primitives, an attacker exploits format strings to override statically stored strings via vulnerable \verb|printf| calls:

\begin{enumerate}
    \item Inject a format string similar to Massidda et al. \cite{massidda2024bringing}: \verb|"A%" + (65535 + i) + "c%2$n%p%p%p%p"|. We extend their technique by demonstrating offset control using \verb|%n| syntax, where \verb|n| specifies the target memory address.
    \item Inject a format string to write specific values by sending characters equal to the target's integer representation (e.g., overriding \verb|b"A"| requires $ord("A") = 41$ characters). Request size limits make long string overwrites nearly impossible in one step; we achieved at most two characters per request.
\end{enumerate}

Constraints are significant: attackers need precise static memory offsets, payload size is limited by server request caps, and character-by-character overwrites make complex queries infeasible.
\par Read primitives are technically simpler but require unconventional architectures. Attackers must observe vulnerable \verb|printf| output to dump sensitive memory segments. In our examples, we assume attackers can read \verb|printf| output called after query execution but before results are freed. An attacker injects format strings with $n$ \verb|%p| placeholders to dump specific memory sections. The primary limitation is that stack-to-data distance can be arbitrarily long, requiring many \verb|%p| directives that servers may reject before reaching the WASM module.

\subsubsection{Use After Free (UAF)}
\label{subsubsec:uaf_sqli}

Use After Free (UAF) vulnerabilities allow the manipulation of heap memory. In our PoC, we demonstrate how an attacker can overwrite a freed query string to perform SQL Injection (SQLi).

In a UAF vulnerability, an attacker exploits the algorithm of the \verb|malloc| function, which tends to reuse recently freed chunks for new memory allocations if the requested size is roughly the same as that of the freed chunk. In the example from our PoC, if the attacker provides a \verb|token| roughly the same size as the freed \verb|query|, the memory allocator may place \verb|user_token| in the exact location of the old \verb|query| variable. When the program subsequently accesses the dangling \verb|query| pointer, it processes the attacker's injected SQL.

The primary constraint for successful exploitation is precision: the injected payload must closely match the size of the freed chunk to trigger the allocator's reuse mechanism.

\subsubsection{Integer Overflow}
\label{subsubsec:iof_sqli}

Integer overflows in WASM often exploit the dissonance between JavaScript's 64-bit floating point integers (safe up to $2^{53}-1$) and C's 32-bit integers. We examined a scenario where a frontend JavaScript check forbids access to an ID of 0. The backend C code, however, accepts the ID as a standard \verb|int|.

If an attacker sends an \verb|user_id| with value $2^{32}-1$, the JavaScript frontend will validate it as non-zero. However, when passed to the WASM module, the value overflows the 32-bit integer limit and wraps around to 0. The backend then unwittingly retrieves the restricted record for ID 0. This vector is particularly dangerous as it requires no memory corruption, purely exploiting data type mismatches between the host (JS) and the guest (WASM).

\subsection{Server Side Template Injection (SSTI)}
\label{sec:resultsssti}

We replicated the architectural approach used for SQL injections to demonstrate how a vulnerable WASM module might lead to the exploitation of Server Side Template Injections (SSTI) vulnerabilities. The exploitation technique was classified as "Chained, not-blind" in chapter \ref{subsec:meth_web}. The environment for our PoCs consists of a NodeJS/Express frontend utilizing the Pug template engine to render dynamic HTML, coupled with a WASM backend.

In our PoC, the WASM module is responsible for generating specific HTML components, such as a \verb|<script>| tag containing arbitrary content. To prevent XSS attacks, the backend WASM module generates a nonce (a one-time random value) that is embedded within the script tag. The frontend, however, trusts this nonce and incorporates it directly into the template without additional validation. If an attacker manipulates the WASM module to return a corrupted nonce containing template syntax (e.g., Pug interpolation like \verb|#{7*7}|), the frontend will execute this injected code during rendering, resulting in Arbitrary Code Execution (ACE). The core vulnerability arises not within the web application itself, but from its implicit trust in the WASM module's output—specifically, the generated nonce. Our PoCs demonstrate three distinct exploitation vectors to achieve this outcome.

\subsubsection{Stack-based Buffer Overflow}
\label{subsubsec:bof_ssti}

This scenario assumes a slight variation of the example proposed in our PoC for SSTI vulnerabilities, in which the WASM function generates a nonce stored on the stack. We crafted a PoC where an unsafe copy operation allows an adjacent buffer to overwrite the nonce variable. Successful exploitation requires a precise overflow of the buffer to overwrite the nonce with a template payload.

By filling the buffer and overflowing into the nonce's memory space, an attacker can replace the legitimate nonce with this payload. When the web application receives the return value and compiles the template, it evaluates the injected syntax (e.g., calculating \(7\times7=49\)), confirming that the binary exploitation successfully propagated to the template engine.

We observed that this vector is the most stable among those tested, confirming similar findings from the SQLi analysis in section \ref{subsubsec:bof_sqli}. The exploitation path is straightforward, with constraints limited to the accurate overwriting of the nonce variable.

\subsubsection{Uncontrolled Format Strings}

Our PoC for the uncontrolled format string vulnerability builds on the same architecture used in the previous section. In this case, the nonce is stored in the binary's static memory section rather than on the stack. Using the same write techniques established in the SQL injection analysis in Section \ref{subsubsec:ufs_sqli}, an attacker can overwrite the static nonce with malicious template syntax. The frontend then renders this manipulated static value, triggering the execution of the injected code.

A crucial constraint identified for this vulnerability is persistence: the target variable must be stored statically (e.g., as a global variable or static local) rather than being re-initialized on every call. Otherwise, the overwrite would be reset before it could be returned to the web application. Exploiting uncontrolled format strings in this context is less stable than stack-based buffer overflows, as it faces the same limitations and exploitation challenges described in Section \ref{subsubsec:ufs_sqli} regarding SQLi.

\subsubsection{Use After Free}

The PoC for the Use-After-Free (UAF) vulnerability also utilizes the same web application architecture outlined previously. In this scenario, the nonce variable is dynamically allocated on the heap. The variable is subsequently freed but remains referenced by a dangling pointer when the function returns, allowing an attacker to reallocate that memory space with a user-controlled payload.

Exploitation requires the following sequence of actions:
\begin{enumerate}
    \item Allocation: The attacker triggers the initial creation of the nonce.
    \item Free and Reallocation: The attacker initiates a logic path where the nonce is freed, immediately followed by the allocation of a new user-controlled string (the payload).
    \item Execution: If the payload is of a size similar to that of the freed nonce, the memory allocator reuses the same chunk. The dangling pointer, intended to reference the nonce, now points to the attacker's template payload.
\end{enumerate}

Consequently, the WASM module returns the attacker's payload to the frontend as if it were the valid nonce, resulting in successful template injection. While viable, this vector exhibited instability during our experiments, akin to the challenges faced in the SQLi context described in Section \ref{subsubsec:uaf_sqli}. The dynamic nature of heap allocations introduces unpredictability, making consistent exploitation more challenging compared to stack-based overflows.

\subsection{XS-Leaks} \label{sec:resultsxsleaks}

Our investigation into XS-Leaks focuses on the exploitation of side channels, specifically timing attacks leveraging Regular Expression Denial of Service (ReDoS) \cite{redos}. Within our classification framework, this represents a "Direct, blind" vulnerability: although the attack interacts directly with the WASM module, the application returns no visible output to the attacker, necessitating inference based on response latency.

The architecture of our PoC consists of a NodeJS/Express frontend interfacing with a backend WASM module compiled from C. The WASM module employs the PCRE2 regex engine \cite{pcre2} and acts as a search service for user-stored secrets based on attacker-supplied regex patterns. While the frontend enforces authentication and input sanitization, the backend contains binary vulnerabilities that can be exploited to manipulate the sanitized regex patterns used in searches.

The attacker's goal is to exfiltrate sensitive user secrets stored within the WASM module's memory. An attacker cannot directly read these secrets (e.g., by accessing the content of the private search results page) due to the Same-Origin Policy (SOP) \cite{sop}, which prevents an external site from reading the content of an HTML page hosted on a different origin. Instead, exploitation is achieved by introducing a side-channel attack through the following actions:
\begin{enumerate}
    \item First, the attacker uses a Cross-Site Request Forgery (CSRF) attack combined with a binary vulnerability to overwrite the legitimate search regex in the WASM memory with a malicious ReDoS vector (e.g., \verb|^secret_prefix(.+){21}|).
    \item Then, the attacker forces the victim's browser to perform a search against their own stored secret using the injected regex.
    \item To infer the page status, the attacker measures the response time. A significant delay indicates that the regex engine attempted a complex match against the injected regex, confirming that the secret begins with the guessed prefix. Conversely, an immediate response indicates a mismatch.
    \item Finally, this process is repeated character-by-character to reconstruct the user's secret.
\end{enumerate}

A universal limitation observed across all vectors is the dependence of regex complexity on the secret length; for instance, a reliable 1-second delay typically requires a secret exceeding 20 characters when using standard ReDoS payloads.

\subsubsection{Stack-based Buffer Overflow}
\label{subsubsec:bof_xs_leaks}

In our PoC for a buffer overflow vulnerability, the WASM module stores user secrets in a global array adjacent to the \verb|search_pattern| variable. Due to WebAssembly's linear memory model, a buffer overflow in the secrets array allows the global search pattern to be directly overwritten.

Since the PCRE2 engine compiles valid (albeit malicious) regexes without error, we successfully injected ReDoS patterns. By measuring the execution time of the \verb|pcre2_match| function, we could reliably confirm the presence of specific characters in the victim's secret, validating the feasibility of this attack vector.

The constraints and limitations of this exploitation path are minimal and consistent with those described in Section \ref{subsubsec:bof_sqli} regarding SQLi. The primary requirement is the precise overwriting of the \verb|search_pattern| variable.

\subsubsection{Uncontrolled Format String}

We attempted to replicate the attack using an uncontrolled format string vulnerability to overwrite the search pattern stored on the stack or in static memory. The theoretical approach involved using the \verb|%n| specifier to write the malicious regex character-by-character.

However, this attempt yielded negative results. During experimentation, the PCRE2 engine consistently panicked or crashed when compiling a regex constructed via format string injection. While the precise cause of this instability remains undetermined, our results suggest that complex memory corruption via format strings may inadvertently corrupt internal structures required by the regex engine, rendering this specific vector ineffective for ReDoS in our environment. We thus consider this vector non-viable for XS-Leak exploitation in our current PoC setup, although further research may be warranted to explore alternative approaches.

\subsubsection{Use After Free (UAF)}

Our UAF PoC leverages a scenario in which the \verb|search_pattern| is dynamically allocated, freed, and subsequently referenced.

The exploitation sequence proceeds as follows:
\begin{enumerate}
    \item The attacker triggers a search that allocates the default pattern.
    \item The attacker triggers a function that frees this pattern.
    \item The attacker immediately provides malicious input (the ReDoS regex), which the allocator places into the newly freed memory chunk.
    \item When the application reuses the dangling pointer to the search pattern, it executes the attacker's ReDoS regex.
\end{enumerate}

We successfully exploited this vulnerability to leak user secrets. However, as with the format string vector, we observed occasional instability where the PCRE2 compilation would fail, likely due to heap metadata corruption during the allocation/deallocation cycles. Despite this, we found the UAF vector to be generally viable for XS-Leak exploitation, albeit with some reliability concerns that merit further investigation.

\subsection{Discussions}
\label{sec:discussions}

Our experiments show that exploitation difficulty varies across vulnerabilities.

Uncontrolled format strings are the hardest without code insight: they require precise target addresses and many implementation-dependent preconditions to make overridden memory readable.

Buffer overflows are relatively easy to exploit: attackers can infer buffer sizes by trial and error without source access, provided inputs avoid crashing the program.

Use-after-free vulnerabilities are also hard to spot without source, but constraints can be bypassed with careful heap shaping.

Overall, binary exploitation tends to destabilize WASM modules; crashes often occur before the bug can be fully leveraged—especially for format strings and UAF.

\subsubsection{Defenses and Mitigations}
\label{sec:defenses}

During our PoCs and analysis, we identified practical defenses for running unsanitized C binaries compiled to WASM in web applications. Some apply at compile time (e.g., flags), others are general best practices:

\begin{itemize}
    \item Treat values crossing the JavaScript–WASM boundary as untrusted. Validate types and ranges on both sides; watch integer width mismatches (Section \ref{subsubsec:iof_sqli}) and canonicalize string lengths/encodings to prevent wrap-around and buffer overflows (Sections \ref{subsubsec:bof_sqli}, \ref{subsubsec:bof_ssti}, \ref{subsubsec:bof_xs_leaks}).
    \item Reduce exported/imported interfaces to the minimum necessary functions and memory regions. This shrinks attack primitives (e.g., arbitrary read/write) and simplifies auditing.
    \item Apply standard web protections. Prepared statements help but are insufficient if query templates are mutable; input sanitization and output encoding remain effective even when the WASM module is vulnerable.
    \item Enable compiler/runtime hardening when speed is not paramount. Emscripten options add stack smashing detection and out-of-bounds checks. Overhead may increase, but exploitation risk drops significantly.
\end{itemize}

A layered approach—interface validation, reduced surface area, compiler/runtime hardening, and runtime detection—reduces both the likelihood and impact of exploitation. In some cases, hardening can nearly eliminate specific vulnerability classes at a performance cost.

\section{Conclusions}
\label{ch:conclusions}

We showed how WASM inherits C‑style vulnerabilities and how attackers can leverage them to trigger web application issues.

Our PoCs demonstrate these flaws reappear in web contexts and can cause unexpected web vulnerabilities. Even without direct interaction between WASM output and the app, timing side‑channels (Section \ref{sec:resultsxsleaks}) can be exploited.

Some vulnerabilities (Buffer Overflows, Use After Free) require little program insight; others (Format Strings) are rarely exploitable in black‑box settings. Even with prepared statements, attackers can still leverage these flaws.

Developers can still apply protections:

\begin{itemize}
  \item \textbf{Sanitize WASM interfaces.} Validate and constrain values crossing the JS–WASM boundary
    before they influence control flow or security‑critical logic.
  \item \textbf{Enable compiler sanitizers.} Use compiler/runtime checks (e.g., Emscripten) for out‑of‑bounds
    access and use‑after‑free.
\end{itemize}

Treat WASM modules with the same security rigor as native applications in web environments, and continue developing robust practices specific to WASM to mitigate these risks.

\section*{Acknowledgment}
This work was partially supported by project SERICS (PE00000014) under the NRRP MUR program funded by the EU - NGEU.

\bibliographystyle{apalike}
{\small
\bibliography{main}}

\end{document}